\documentclass[,final]{aipproc}
\layoutstyle{8x11double}
\usepackage{aas_macros}

\def\xmm {\emph{XMM-Newton}}
\def\cxo {\emph{Chandra}}
\def\suz {\emph{Suzaku}}

\def\xte {\emph{RXTE}}
\def\int {\emph{INTEGRAL}}
\def\asca {\emph{ASCA}}
\def\src {SGR\,1806$-$20}

\def\flux {\mbox{erg cm$^{-2}$ s$^{-1}$}}
\def\lum {\mbox{erg s$^{-1}$}}
\def\nh {$N_{\rm H}$}

\begin{document}
\title{The first \suz\ observation of \src}

\author{P.~Esposito}{address={University of Pavia, Department of Nuclear and Theoretical Physics and INFN-Pavia, Pavia, Italy} ,altaddress={INAF - Istituto di Astrofisica Spaziale e Fisica Cosmica Milano, Milano, Italy}}

\author{A.~Tiengo}{address={INAF - Istituto di Astrofisica Spaziale e Fisica Cosmica Milano, Milano, Italy}}

\author{S.~Zane}{address={Mullard Space Science Laboratory, University College London, Holmbury St. Mary, UK}}

\author{R.~Turolla}{address={University of Padua, Department of Physics, Padova, Italy}
 ,altaddress={Mullard Space Science Laboratory, University College London, Holmbury St. Mary, UK}}

\author{S.~Mereghetti}{address={INAF - Istituto di Astrofisica Spaziale e Fisica Cosmica Milano, Milano, Italy}}

\author{D.~G\"{o}tz}{address={CEA Saclay, DSM/DAPNIA/Service d'Astrophysique, Gif-sur-Yvette, France}}

\author{G.~L.~Israel}{address={INAF - Osservatorio Astronomico di Roma, Monteporzio Catone, Italy}}

\author{N.~Rea}{address={University of Amsterdam, Astronomical Institute ``Anton Pannekoek'', Amsterdam, The Netherlands}}

\begin{abstract}
The soft gamma-ray repeater \src\ has been attracting a lot of attention owing to the fact that in December 2004 it emitted the most powerful giant flare ever observed. Here we present the results of the first \suz\ observation of \src, that seems to have reached a state characterized by a flux close to the pre-flare level and by a relatively soft spectrum. Despite this, the source remained quite active, as testified by several short bursts observed by \suz. We discuss the broadband spectral properties of \src\ in the context  of the magnetar model, considering its recent theoretical developments.
\end{abstract}

\maketitle


\section{Introduction}

The Soft Gamma-ray Repeater (SGR) \src\ was discovered in 1979 as a transient source of high-energy photons \citep{laros86}. SGRs emit sporadic and short (\mbox{$\sim$0.1 s}) bursts of (relatively) soft gamma-rays with luminosity of $10^{40}$--$10^{41}$ \lum during periods of activity, that are often broken by long intervals of quiescence. 
SGRs are thought to be magnetars: highly magnetized neutron stars with field strength of \mbox{$10^{14}$--$10^{15}$ G}, considerably larger than those of the majority of radio pulsars \citep[][]{thompson95,thompson96}.
In the magnetar framework, the ultimate source of energy for the bursts and the quiescent emission is the ultra-strong magnetic field. 
The persistent X-ray counterpart of \src\ was observed for the first time with the \asca\ satellite in 1993 \citep{murakami94} in soft X-rays (\mbox{$<$10 keV}) and more recently also in the hard X-ray range with \int\ \citep{mgm05}. 
A \xte\ observation led to the discovery of pulsations in the persistent emission with period $P\simeq7.47$ s and period derivative \mbox{$\dot{P}\simeq2.6\times10^{-3}$ s  yr$^{-1}$} \citep{kouveliotou98}. 
Under the assumption of pure magnetic dipole braking, these values imply a surface magnetic field strength of \mbox{$8\times10^{14}$ G}, strongly supporting the magnetar model.\\
\indent Both the burst rate and the X-ray persistent emission of \src\ started increasing during 2003 and throughout 2004 \citep{mte05}, culminating with the giant flare of 2004 December 27, during which \mbox{$\sim$$10^{47}$ erg} were released (assuming isotropic luminosity and a distance of 15 kpc), $\sim$100 times more than in the flares observed from SGR\,0526--66 in 1979 and from SGR\,1900+14 in 1998 \citep{hurley05}.
After the giant flare, the persistent X-ray flux of \src\ started to decrease, eventually recovering the ``historical'' pre-flare level, and its X-ray spectrum to soften \citep{met07}. 
A similar flux decrease have been observed from its hard X-ray and infrared counterparts \citep{met07}.
Here we present the results of the first \suz\ observation of \src\ and we discuss the spectral properties of its persistent emission. 

\section{Observation and data analysis}
\begin{figure}
  \includegraphics[width=.3\textwidth,angle=-90]{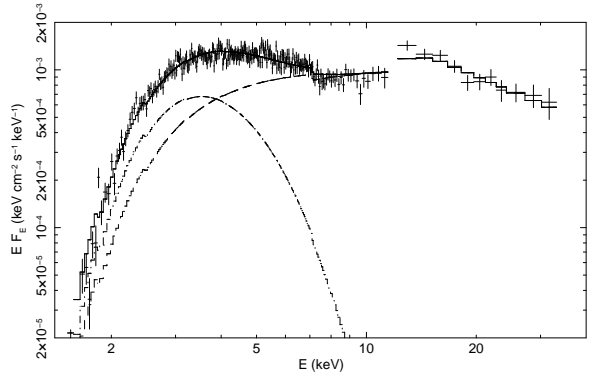}
  \caption{Broad-band \suz\ spectrum of \src. The \suz\ XIS and HXD-PIN data are fit with the broken power-law (dashed line) plus blackbody  (dot-dashed line) model. The XIS\,1 data are not shown for clarity.}
  \label{suz-broad}
\end{figure}

The \suz\ X-ray observatory \citep{mitsuda07} carries on board the XIS spectrometers operating in the \mbox{0.2--12 keV} energy band, and the HXD collimated detector, which covers the \mbox{10--70 keV} energy range with PIN diodes and the \mbox{40--600 keV} with GSO scintillators.
The observation of \src\ started on  2006 September 09 at \mbox{23:13 UT}, and ended on September 11 at 04:01 UT, for a net exposure of \mbox{46.4 ks} in the XIS and 48.4 ks in the PIN, after a screening based on standard criteria \citep[see][for more details]{emt07}.
By an inspection of the \suz\ lightcurves we found several bursts, that were excluded by appropriate
time selections to derive the spectral results reported below (the analysis of the bursts is presented in \citep{emt07}).
Apart from the bursts, no variability in the XIS light curves of \src\ was detected (the 8 s time resolution of the XIS does not allow to detect the pulsation).\\
\indent Owing to the high interstellar absorption, very few counts were detected from \src\ at low energies and thus we limited the spectral analysis to the \mbox{1.5--12 keV} energy range.
We fit simultaneously the four XIS spectra adopting a power-law and a power-law plus blackbody model.
The reduced $\chi^2$  of the former fit, \mbox{$\chi^2_r = 1.16$} for 283 degrees of freedom (d.o.f.), corresponding to a null hypothesis probability (n.h.p.) of 0.03, is not completely satisfactory. The power-law plus blackbody model provided a better fit, with \mbox{$\chi^2_r = 0.98$} for \mbox{281 d.o.f.} (n.h.p. = 0.6). 
The best fit parameters are reported in Table \ref{fit} for both models. 
The presence of the blackbody component is consistent with the findings of several \xmm\ observations \citep{mte05,met07}.
\begin{table}
\begin{tabular}{ccc}
\hline
Parameter &  \multicolumn{2}{c}{Value}\\
 & PL & PL + BB \\
\hline
\nh\ (10$^{22}$ cm$^{-2}$) & $6.4\pm0.2$ & $7.1^{+0.6}_{-0.5}$ \\
$\Gamma$ & $2.03\pm0.04$ & $1.8\pm0.1$ \\
$k_BT$ (keV) & -- & $0.49^{+0.08}_{-0.07}$ \\
R$_{\rm{BB}}$\tablenote{Radius at infinity assuming a distance of 15 kpc.}\ \  (km) & -- & $5^{+9}_{-2}$ \\
Flux\tablenote{Flux in the 2--10 keV range, corrected for the absorption.}\ \ \ ($10^{-11}$\flux) & $1.90\pm0.04$ & $2.1\pm0.1$ \\
$\chi^{2}_{r}$ (d.o.f.) & 1.16 (283) & 0.98 (281) \\
\hline
\end{tabular}
\caption{Summary of the spectral results in the \mbox{1.5--12 keV} energy range. Errors are quoted at the 90\% confidence level for a single interesting parameter.}
\label{fit}
\end{table}\\
\indent The advantages of HXD over previous non imaging instruments are its small field of view ($34'$$\times$$34'$ FWHM below $\sim$100 keV) and a low instrumental background. 
The images obtained from \int\ very deep exposures do not show contaminating point sources within the HXD field of view \citep[see][Fig.\,1]{mgm05} and no bright and hard X-ray point sources below 10 keV have been found in the SIMBAD database. 
However, given that \src\ lies at low Galactic latitude and longitude ($b\simeq0^{\circ}$ and $l\simeq10^{\circ}$), the study of its emission in the hard X\,/\,soft gamma-ray band is complicated by the presence of the diffuse emission from the Galactic Ridge \citep[GR; e.g.,][]{valinia98}. A flux possibly associated with \src\ is detected in the HXD-PIN data up to $\sim$40 keV (no significant emission is detected in the GSO data). 
The instrumental background counts obtained by simulations based on the present knowledge of PIN in-orbit performances (instrumental background events were provided by the HXD Team), are about 70\% of the $\sim$26,400 total counts in the 12--40 keV band.
We note that to account for the whole signal detected in the HXD-PIN instrument, the GR emission in the HXD field of view should be $\sim$7 times higher than that reported by \citep{valinia98}. This seems very unlikely and therefore we consider significant the detection of \src. However, both its spectral shape and flux are subject to the uncertainty reflecting the coarse knowledge of the GR contribution to the background.
Including the cosmic diffuse and GR emission as fixed components \citep[see][for details]{emt07}, we fitted the PIN spectrum to a  power-law model. The best-fit  parameters (\mbox{$\chi^2_r=0.90$} for 10 d.o.f.) are $\Gamma=2.0\pm0.2$ and flux in the \mbox{20--60 keV} range of \mbox{$(3.0\pm0.5)\times10^{-11}$ \flux}. 
Fitting together the PIN and XIS\ spectra (for the broadband analysis we added the spectra of the three front-illuminated  CCDs: XIS\,0, 2, and 3), we find that the PIN data must be scaled downward by a factor of $\sim$2 to be consistent with the parameters derived in the 1.5--12 keV energy range.  
This scaling factor is unacceptably large, since the uncertainty in the relative calibration of the two instruments in the energy band considered here is less than 20\% \citep{kokubun07}. 
To better reproduce the broadband spectrum we tried a broken power-law plus blackbody model, with a normalization factor between the instruments kept at $<$1.2. We find an acceptable fit (\mbox{$\chi^2_r = 1.09$} for \mbox{354 d.o.f.}; \mbox{n.h.p. = 0.13}) with the photon index changing from \mbox{$1.0\pm0.1$} below the break at \mbox{$16\pm2$ keV} to $2.2^{+0.4}_{-0.2}$ above it, $k_B T=0.8\pm0.1$ keV, \mbox{$R_{\rm{BB}}= 2.5^{+0.4}_{-0.3}$ km} (at \mbox{15 kpc}), and \mbox{$N_{\rm H}=5.6^{+0.3}_{-0.4}\times10^{22}$ $\rm cm^{-2}$} (see Fig.\,\ref{suz-broad}). The corresponding 2--10 keV and 20--60 keV unabsorbed fluxes are \mbox{$\sim$$2 \times 10^{-11}$ \flux} and \mbox{$\sim$$3 \times 10^{-11}$ \flux}, respectively.

\section{Discussion}
Although the  \suz/HXD does not have imaging capabilities, we know thanks to  \int\ observations that no other bright hard X-ray point sources are present in its field of view.  
The uncertainties in the instrumental background (currently at the $\sim$5\% level, \citep{kokubun07}) and in the modelling of the GR emission are a more concern.
Future improvements in the knowledge of these components may eventually allow us to obtain more robust conclusions. 
With all these caveats in mind we proceed now to discuss the broadband spectrum of \src.\\
\indent
With respect to the XIS, the HXD data show an ``excess'' (see Fig.\,\ref{suz-broad}) that cannot be completely ascribed to calibration uncertainties between the instruments. 
Given the lack of a direct measure of the GR emission around  \src, we cannot exclude that this excess is due an underestimation of such contribution to the background. If instead it is a real feature of the source spectrum, its broadband spectrum could be empirically modeled adopting a power-law with the photon-index changing from $\sim$1 to $\sim$2 at \mbox{$\sim$16 keV}, and a blackbody component with \mbox{$k_BT\sim0.8$ keV}. This would agree with the results reported by \citep{gotz06}, who point out that the hard tails of the SGRs are softer than the power-law components measured below \mbox{10 keV}.
If confirmed, the presence of down-break in the 10--20 keV
spectrum of \src\ has remarkable physical implications. 
The HXD data suggests that the soft power-law can extend well above the \xmm\ and \cxo\ spectral band, potentially up to $\sim$15--20~keV, before it turns downwards to match the steeper power-law spectrum measured at higher energy.
Up to now, it is still unclear wether a single physical mechanism is responsible for the observed spectrum in the \mbox{1--200 keV} band, or the soft ($\sim$1--10 keV) and the hard ($>$20 keV) emission are produced in different ways. In the latter case, the overall spectral shape changes in time and the difference between the sources may be produced by the different relative strength of the soft and hard components.\\
\indent Recently, theoretical computations of the magnetars broadband spectra have been presented, in the attempt to identify a common physical mechanism which could explain both the hard and soft X-ray emission. 
As suggested by \citep{tlk02}, within the twisted magnetosphere model a substantial non-thermal (power-law) component appears as the result of resonant cyclotron up-scattering of soft surface-emitted photons onto charges flowing into the magnetosphere. 
A simple one-dimensional treatment of resonant compton scattering (RCS) has been presented by
\citep{lyutikov06} and more detailed, 3-dimensional computations have been performed by \citep{fernandez07}. 
Since all these treatments are based on the use of the non-relativistic (Thompson) magnetic cross section, spectral predictions are supposedly accurate only up to $\sim$30--50~keV. This is, however, well above the spectral break detected by \suz. 
Interestingly, some of the model spectra presented by \citep{fernandez07} exhibit a downward break in the tens of keV range and have an overall shape quite reminiscent of the XIS/HXD spectrum of \src. 
In particular (see their Fig.\,6 and Fig.\,11), when assuming a (non-thermal) top-hat velocity distribution or a broadband velocity distribution for the magnetospheric charges, multiple peaks can appear in the spectrum. 
Due to the complexity of the model, it is difficult to predict the relative position of the peaks, but it may well be that the downturn detected in our data witnesses the presence of a second ``hump'' (in addition to the main thermal one) in the range \mbox{10--20~keV}. 
Nobili, Turolla, \& Zane (in preparation), assuming a \mbox{1-dimensional} thermal electron distribution superimposed to a (constant) bulk velocity, found also double humped spectra. 
In this case the second (and only) hump occurs when resonant scattering is efficient enough to fill the Wien peak at the temperature of the comptonizing particles. 
A spectral break at $\sim$15~keV would translate then in a temperature of $\sim$5 keV for the magnetospheric electrons. 
Therefore, if  confirmed, the spectral break in the X-ray data of \src\ may lend further support to the RCS model and prove useful in constraining the physical parameters of the model.


\begin{theacknowledgments}
PE thanks the McGill University and in particular V.~M.~Kaspi, A.~Cumming, M.~Zamfir, R.~P.~Breton, and J.~Gardner, for the warm hospitality.
\end{theacknowledgments}

\bibliographystyle{aipproc}   
\bibliography{biblio}

\end{document}